\documentstyle[aps,xspace]{revtex}

\begin{document}

\newcommand{\kn}{${\rm K}^0$\xspace}
\newcommand{\knt}{$\widetilde{{\rm K}^0}$\xspace}

\newcommand{\nri}{$\nu_{\rm I}$\xspace}
\newcommand{\nrii}{$\nu_{\rm II}$\xspace}
\newcommand{\nrit}{$\tilde{\nu}_{\rm I}$\xspace}
\newcommand{\nriit}{$\tilde{\nu}_{\rm II}$\xspace}


\title{Neutrino Flavor Tagging in a Four-Neutrino Mixing and Oscillation 
  Model}

\author{E.M. Lipmanov\footnote{e-mail: elipmanov@yahoo.com}}

\address{40 Wallingford Rd. \#272, Brighton, MA 02135, USA}

\maketitle
 
\begin{abstract}
  A neutrino mass dominance quantity is introduced for tagging the
  neutrino flavor in the phenomenological two-parameter four neutrino
  mixing matrix with two neutrino mass doublets and thorough maximal
  neutrino doublet mixing.  While there is no hierarchy of the neutrino
  masses in the neutrino flavor eigenstates of this model, it may rather
  be a special hierarchy of the mass dominance ratios in these
  eigenstates.  A neutrino flavor hierarchy condition is suggested: a
  direct link between the neutrino flavor and the flavor of the charged
  leptons which interconnects the two mixing angles, $\theta$ and
  $\phi$, via the charged lepton mass ratios, with the net result
  $tg^2\phi = (tg^2\theta)^\gamma, \gamma \cong 2.06$.  It leads to
  distinct inferences testable at SNO and Super-K.
\end{abstract}






\vspace{5mm}

\noindent
{\bf keywords:} {\it two mass doublets; neutrino flavor tagging}

\vspace{5mm}


In the SM, the flavor quantity $f_l$ of a charged lepton $l$ can be
associated definitely only with its mass value $m_l$,
\begin{equation}
  \label{eq:fi}
  f_l\equiv m_l
\end{equation}
\noindent
In the three-neutrino mixing model with the see-saw neutrino mass
hierarchy mechanism~\cite{seesaw}, the flavor of the weak interaction
neutrino eigenstate $\nu_l$ can be tagged by its dominant mass
eigenstate (as in the quark case).  At first sight it seems that the
tagging of the neutrino flavor gets entangled excessively in the
neutrino flavor compositions of the four-neutrino mixing model with two
nearly mass-degenerate neutrino doublets~\cite{numodels,bilenky-rev}.
In fact, as shown in this note, it is not necessarily so.  In the
phenomenological four-neutrino mixing matrix with a simple doublet
neutrino symmetry~\cite{lipmanov1} and an extended
\kn-analogy~\cite{lipmanov2}, with only two ($\theta$ and $\phi$) mixing
angles, which is in tune with the majority of the neutrino oscillation
data, the three flavor neutrinos $\nu_e$, $\nu_\mu$, and $\nu_\tau$,
plus one sterile neutrino $\nu_{st}$, are
\begin{eqnarray}
  \label{eq:nue}
  \nu_e    & = & (\nu_1^s\cos\theta + \nu_2^s\sin\theta)_L\\
  \label{eq:numu}
  \nu_\mu  & = & (-\nu_1^s\sin\theta + \nu_2^s\cos\theta)_L\\
  \label{eq:nutau}
  \nu_\tau & = & (\nu_1^a\sin\phi + \nu_2^a\cos\phi)_L\\
  \label{eq:nus}
  \nu_{st} & = & (\nu_1^a\cos\phi - \nu_2^a\sin\phi)_L
\end{eqnarray}
\noindent
with the notations
\begin{equation}
  \label{eq:ni_il}
  \nu_{iL}^s = \frac{1}{\sqrt{2}}(\nu_i+\nu_i^\prime)_L,~~~ 
  \nu_{iL}^a = \frac{1}{\sqrt{2}}(\nu_i-\nu_i^\prime)_L,
\end{equation}
\noindent
$i=1,2$ and ($\nu_1,\nu_1^\prime$), ($\nu_2,\nu_2^\prime$) are four
Majorana neutrino mass eigenstates grouped in two mass doublets with
opposite CP-parities in each of the two doublets.  The
data of the solar, atmospheric, and LSND experiments can be explained 
with the following values of the neutrino
mass squared differences~\cite{numodels,bilenky-rev}:
\begin{eqnarray}
  \label{eq:dm2}
  \Delta m_1^2 & = & m_1^2-m_1^{\prime\, 2} = \Delta m^2_{solar} \approx
  10^{-10}(Vac),~{\rm or} \approx 10^{-5}(MSW)~{\rm eV}^2,\nonumber\\
  \Delta m_2^2 & = & m_2^2-m_2^{\prime\, 2} = \Delta m^2_{atm} \approx
  10^{-3} {\Large -} 10^{-2}~{\rm eV}^2,\\
  \Delta m_{12}^2 & \cong & m_2^2-m_1^2 \approx 1~{\rm eV}^2,\nonumber
\end{eqnarray}
\noindent
with a possible scheme of the neutrino mass spectrum
\renewcommand{\theequation}{8A}
\begin{equation}
  \label{eq:doublet}
  \underbrace{\overbrace{m_1 < m_1^\prime}^{solar} \ll 
    \overbrace{m_2 < m_2^\prime}^{atm}}_{LSND}~.
\end{equation}
\renewcommand{\theequation}{\arabic{equation}}\setcounter{equation}{8}
\noindent
In the other possible scheme~(8B) the positions of the
``solar'' and ``atm'' doublet splittings are interchanged.  

There are two entirely different kinds of neutrino mixings in this
model: 1).~The thorough maximal doublet neutrino mixings which do not
introduce any free parameters here; and 2).~The apparently small mixings
between different neutrino doublets with mixing angles $\theta$ and
$\phi$ which remained free parameters.  The origin of the same neutrino
doublet maximal mixings is likely related to the special feature of the
Majorana neutrino mass physics without any visible connection to the
lepton flavor problem; no such connection is also seen in the mass
expectation values (arithmetic mean masses, in this case) of the two
pairs of the auxiliary neutrino fields ($\nu_1^s, \nu_1^a$) and
($\nu_2^s, \nu_2^a$) in the superpositions~(\ref{eq:nue}),
(\ref{eq:numu}), and ({\ref{eq:nutau}), they are equal within each
pair.  In the limit of mass-degenerate Majorana neutrino doublets
(comp. Eq.~(\ref{eq:dm2})), the two auxiliary neutrino fields $\nu_1^s$
and $\nu_2^s$ become 4-component Dirac neutrinos ($\nu_1^a$ and
$\nu_2^a$ -- the antineutrinos) carrying lepton charge, which is
conserved in the the lepton weak interactions~\cite{lipmanov2}.  
  On the contrary, the origin of the small mixings of the different mass
  doublet neutrinos must be connected with the lepton flavor problem
  because, just by the definition, these mixings do shape the neutrino
  weak interaction eigenstates~(\ref{eq:nue}), (\ref{eq:numu}) and
  (\ref{eq:nutau}), which differ from each other, in essence, only by
  their mixing amplitudes.  The only physical quantity which can tag the
  flavor of an individual neutrino composition in the equations
  (\ref{eq:nue}), (\ref{eq:numu}) and (\ref{eq:nutau}) is the relative
  probability of its auxiliary neutrino fields $\nu_1^s$ and $\nu_2^s$,
  or $\nu_1^a$ and $\nu_2^a$.  Consequently, a new neutrino flavor
  concept can be incorporated in the present model: the right
  identification of the neutrino weak interaction eigenstates
  $\nu_l,l=e, \mu$ and $\tau$, is determined by the neutrino flavor
  quantities $f_{\nu_{l}}$,
\begin{equation}
  \label{eq:f_nu_l}
  f_{\nu_{l}} \equiv \left [ W(\nu_{2;1}^{s(a)})/W(\nu_{1;2}^{s(a)}) 
  \right ]_{\nu_{l}},
\end{equation}
\noindent
where the two versions separated by semicolons are for the two neutrino
mass schemes~(8A) and (8B), respectively.
The quantity $[W(\nu_i^{s(a)}]_{\nu_{l}}, i=1,2$, is the probability of
finding the state $\nu_i^s$ (or $\nu_i^a$) in the neutrino weak
interaction eigenstates $\nu_e$, $\nu_\mu$ and $\nu_\tau$ in
Eqs.~(\ref{eq:nue}), (\ref{eq:numu}) and (\ref{eq:nutau}).  The neutrino
flavor quantities $f_{\nu_{l}}$ in Eq.~(\ref{eq:f_nu_l}) can be
considered as neutrino mass dominance ratios in the neutrino flavor
eigenstates (in the mass scheme~(8B) the mass dominances
are reversed).  The neutrino flavor hierarchy condition gets the form:
\begin{equation}
  \label{eq:f_nu_l-k}
  f_{\nu_{l}} = k(f_l)^r,~~~l=e, \mu, \tau,
\end{equation}
\noindent
where $k$ and $r > 0$ are independent of $l$ constants.  From
Eqs.~(\ref{eq:f_nu_l-k}), (\ref{eq:f_nu_l}) and
(\ref{eq:nue})-(\ref{eq:nutau}) we obtain three algebraic equations,
\begin{eqnarray}
  \label{eq:tg2arr}
  tg^2\theta &~;~& ctg^2\theta = k(m_e)^r,\nonumber\\
  ctg^2\theta &~;~& tg^2\theta = k(m_\mu)^r,\nonumber\\
  ctg^2\phi &~;~& tg^2\phi = k(m_\tau)^r,
\end{eqnarray}
\noindent
for four unknowns $k, r, tg^2\theta$, and $tg^2\phi$.  The solution is
\begin{equation}
  \label{eq:tg2}
  tg^2\theta ~;~ ctg^2\theta=(m_e/m_\mu)^{r/2},
\end{equation}
\noindent
\begin{equation}
  \label{eq:tg}
  tg^2\phi=(tg^2\theta)^\gamma ~,~ \gamma=\ln(m_e m_\mu/m_\tau^2) / 
  \ln(m_e/m_\mu) \cong 2.06.
\end{equation}
\noindent
Equation~(\ref{eq:tg}) is a relation between the two neutrino mixing
angles $\theta$ and $\phi$, independent of the value of the exponent $r$ 
in the main statement~(\ref{eq:f_nu_l-k}).  It ensures that the mixings
between the neutrinos from different doublets are ``small'',
\begin{equation}
  \label{eq:tgphi}
  tg^2\phi = (tg^2\theta)^\gamma \cong 3\times 10^{-7},
\end{equation}
\noindent
if the LSND data, $\sin^2 2\theta \cong 3\times 10^{-3}$~\cite{lsnd},
are accepted ($r\cong 2.7$).  The result~(\ref{eq:tgphi}) leads to 
main inferences: 1).~The amplitudes of the short-baseline
$\nu_\tau\rightarrow\nu_{st}$ transformation and $\nu_\tau$
disappearance oscillations should be $\sin^2 2\phi~
\cong~10^{-6}$, {\it i.e.} much smaller than the LSND oscillation
amplitude; if there are sterile neutrinos of yet unknown cosmic origin,
one could hope to observe them (in the present framework -- almost
exclusively) via the large amplitude long-baseline oscillations
$\nu_{st}\rightarrow\nu_e$ by the resulting electrons.  
2).~A strong $\nu_\mu\rightarrow\nu_\tau$ dominance in the
atmospheric $\nu_\mu$ oscillations ($\nu_e\rightarrow\nu_{st}$ dominance in
the solar $\nu_e$-oscillations).  These inferences are independent of
the choice of the neutrino mass scheme~(8A) or
(8B).  They agree well with the implications of the
standard BBN constraints on the four neutrino mixing phenomenology
discussed in ref.~\cite{bilenky2} and will be tested in the measurements
of the ratios CC/NC in the neutrino experiments such as SNO and
Super-K~\cite{altmodels}.

I would like to thank Alec Habig for interesting conversations.

\end{document}